%% file: main.tex
\IfFileExists{IEEEtran.cls}{%
  \documentclass[lettersize,journal]{IEEEtran}%
}{%
  \documentclass[10pt,twocolumn]{article}%
  \usepackage[margin=0.75in]{geometry}%
}
\usepackage[T1]{fontenc}
\usepackage{amsmath}
\usepackage{amssymb}
\usepackage{graphicx}
\usepackage{booktabs}
\usepackage{array}
\usepackage[table]{xcolor}
\usepackage{tikz}
\usetikzlibrary{arrows.meta,positioning,fit,calc}
\usepackage[hidelinks]{hyperref}
\usepackage{xurl}
\usepackage{cite}
\input{values.tex}
\input{values_v3.tex}
\input{values_v4.tex}
\newcommand{\code}[1]{\texttt{\small #1}}
\newcolumntype{L}[1]{>{\raggedright\arraybackslash}p{#1}}
\definecolor{hdr}{HTML}{1F4E79}
\definecolor{stripe}{HTML}{EAF2FB}
\newcommand{\no}[1]{\textcolor{red!70!black}{#1}}
\newcommand{\yes}[1]{\textcolor{green!45!black}{\textbf{#1}}}
\makeatletter
\newcommand{\ifieee}[1]{\@ifclassloaded{IEEEtran}{#1}{}}
\makeatother

\begin{document}

\title{Measuring GPU Memory Traffic on the NVIDIA GB10 Without C2C Telemetry}
\author{Deepak Panigrahy$^{*}$, Sirisha Kota$^{*}$, Aakash Tyagi, David Kebo, and Rabi Mahapatra%
\thanks{$^{*}$D. Panigrahy and S. Kota contributed equally to this work.}%
\thanks{D. Panigrahy, A. Tyagi, D. Kebo and R. Mahapatra are with the Department of Computer Science and Engineering, Texas A\&M University, College Station, TX 77843 USA (e-mail: deepakp@tamu.edu; tyagi@cse.tamu.edu; davidkebo@tamu.edu; rabi@tamu.edu).}%
\thanks{S. Kota is with The Himalayan Foundation Research Division, Bentonville, AR 72712 USA (e-mail: skota@thehimalayanfoundation.in).}}
\ifieee{\markboth{}{Panigrahy \MakeLowercase{\textit{et al.}}: Measuring GPU Memory Traffic on the GB10 Without C2C Telemetry}}
\maketitle

\begin{abstract}
The NVIDIA GB10, used in the DGX Spark and in desktop AI systems from several vendors, joins an Arm CPU die and a Blackwell GPU die with NVLink C2C.
The GPU has no local memory, so every GPU access that misses in its L2 cache crosses C2C to the LPDDR5X attached to the CPU die.
Yet the software on the systems we tested exposes no counter for this traffic.
We show that five L2 sector counters of the GPU, sampled every 10\,ms through NVIDIA's CUPTI sampling interface by a process that does not attach to the workload, measure GPU memory traffic live and separate copy engine traffic from the read misses and writes of GPU kernels.
For large streaming transfers of known size, the counters agree within \val{ver.gn.maxerr}\% on an Acer GN100 and within 0.005\% on a DGX Spark, track the volume moved regardless of how it is split into copies, and exclude data reused in L2.
The GPU's SMMU counter, a natural alternative, counts address translations rather than bytes.
Applied to an unmodified server, the method shows Mistral 7B decode reading \val{llm.decode_gib_tok}\,GiB per token, the weights it uses plus its key value (KV) cache, at 83\%  of the memory bandwidth NVIDIA specifies.
We release the method as an open source tool.
\end{abstract}
\ifieee{}

\section{Introduction}
\textbf{Unified memory on the GB10.}
The NVIDIA GB10 is used in the DGX Spark, and in the desktop AI systems from several vendors like Acer, ASUS, Dell, GIGABYTE, HP, Lenovo and MSI.
It joins an Arm CPU die and a Blackwell GPU die with NVLink chip to chip (C2C), and both use \val{plat.memory_gb}\,GB of LPDDR5X attached to the CPU die~\cite{hotchips-gb10}.
The GPU has no memory of its own, so GPU accesses that reach memory, rather than being served by the GPU's L2 cache, cross C2C (Fig.~\ref{fig:system}).
How many bytes a workload moves decides whether it is limited by bandwidth, how much energy moving data costs, and how much it competes with the CPU for the same memory.
On a discrete GPU these questions split cleanly: device memory traffic stays on the board, and only explicit transfers cross PCIe.
On the GB10 there is one memory and one path to it, so every kernel load that misses in the GPU's L2 is also C2C traffic.

\textbf{Interconnect telemetry.}
The GB10 systems we tested expose no counter for this traffic.
Grace servers have Linux counters for the bytes crossing C2C~\cite{linux-nv-pmu}, and data center GPUs report interconnect traffic through the Data Center GPU Manager (DCGM), but neither is available on the GB10: DCGM fails to load on both systems we tested (Table~\ref{tab:tools}).

\textbf{Why existing tools fall short.}
Existing tools answer neighboring questions.
Nsight Compute reports the memory traffic of individual kernels, but it replays each kernel several times to collect its counters~\cite{ncu-guide}.
Replay serializes the program, slows it by a large factor, and runs each kernel alone rather than alongside the work it normally overlaps with, so a running server cannot be observed in its normal state.
Nsight Compute also does not see copies or memsets that the driver performs without a kernel.
Nsight Systems records when kernels and copies run, and the size of each explicit copy, which the program states in the call.
A kernel, in contrast, announces no size: how many bytes it reads and writes while it runs (purple and green paths in Fig.~\ref{fig:system}) is known only to hardware counters, and in LLM inference these kernel reads and writes are nearly all of the traffic, so for a running model Nsight Systems shows when kernels run but almost none of the bytes they move.
On the GB10, its sampled GPU metrics report clocks and how busy the streaming multiprocessors (SMs), tensor cores and copy engines are, but no memory or link traffic.
The memory utilization reported by \code{nvidia-smi} is the fraction of time memory was busy, not a volume.
What is missing is a byte count for the whole GPU, sampled during normal execution, without kernel replay.

\textbf{Approach and contributions.}
We find that the gap can be narrowed from the GPU side.
Every GPU request to memory passes the GPU's L2 cache, and the L2 counts the 32\,B sectors it handles, by the unit that made the request.
This letter makes three contributions.
First, we identify five L2 counters that give the GPU's memory traffic, split into copies and kernel reads and writes, and show that they can be sampled live from a separate process.
Second, we validate them against known volumes on two GB10 systems, and show why the GPU's SMMU counter, a natural first candidate, does not measure bytes.
Third, we show what the counters reveal about a real workload: prefill and decode of an LLM server observed without rebuilding or instrumenting it.
Throughout, we call the quantity we measure \emph{GPU memory traffic observed at L2}; under the GB10's topology it estimates the payload carried over C2C, not the bytes on the wire (Sec.~\ref{sec:limits}).

\begin{figure}[t]
\centering
\input{figures/fig1_system.tex}
\caption{GB10: GPU accesses that miss in L2 cross NVLink C2C, for which no counter is exposed.
Every request passes the GPU's L2 cache, whose counters (blue tags) count the sectors of kernel read misses, kernel writes and copy engine traffic.
Our software, a separate process on the Arm cores (dashed band), samples them every 10\,ms; everything above it is hardware.
Die layout after~\cite{hotchips-gb10}.}
\label{fig:system}
\end{figure}
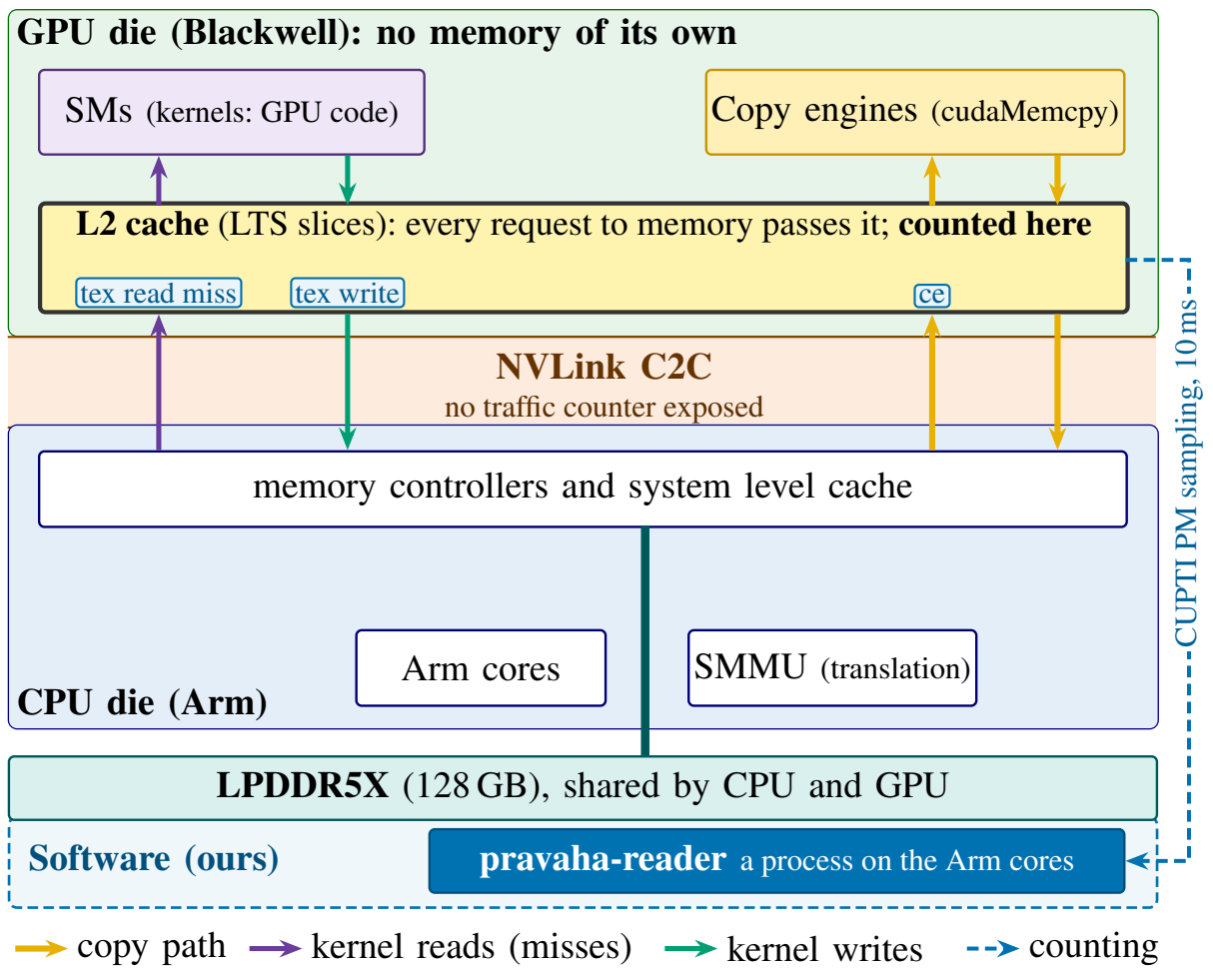

\section{Counter availability and selection}
\textbf{Direct telemetry.}
Table~\ref{tab:tools} lists the interfaces we checked on both systems.
DCGM profiling fails to load, the NVIDIA Management Library (NVML) lists the C2C links but has no query for their traffic, Nsight Compute's C2C section collects no metrics, and the Linux driver for C2C counters finds none, because the firmware lacks the Arm Performance Monitoring Table (APMT) that would locate them.
On Grace, the same driver exposes the C2C and memory counters as \code{perf} units that count bytes per link direction~\cite{linux-nv-pmu}; on the GB10, \code{perf list} shows none of them.
Both systems, from two vendors, ran NVIDIA driver \val{gn.driver} and gave the same result.
We therefore claim that the tested software exposes no C2C traffic counter, not that the chip lacks one.

\begin{table}[t]
\centering
\caption{Interfaces checked for GPU memory traffic on both systems.}
\label{tab:tools}
\footnotesize
\setlength{\tabcolsep}{4pt}
\renewcommand{\arraystretch}{1.05}
\arrayrulecolor{blue!30}
\rowcolors{2}{white}{stripe}
\begin{tabular}{|L{0.40\columnwidth}|L{0.52\columnwidth}|}
\hline
\rowcolor{hdr}
\textcolor{white}{\textbf{Interface}} & \textcolor{white}{\textbf{Result}} \\ \hline
DCGM profiling & \no{fails to load} \\ \hline
NVML & \no{C2C links listed; no traffic query} \\ \hline
Nsight Compute C2C, NVLink & \no{no metrics; 0 links} \\ \hline
Nsight Systems GPU metrics & \no{activity \% only; no traffic} \\ \hline
Linux \code{nvidia\_cspmu}, APMT & \no{no C2C counter units; table absent} \\ \hline
GPU SMMU PMCG (\code{perf}) & \no{counts translations, not bytes} \\ \hline
L2 counters, CUPTI PM sampling & \yes{sectors, live (used)} \\ \hline
\end{tabular}
\end{table}

\textbf{SMMU counters.}\label{sec:smmu}
The System Memory Management Unit (SMMU) translates the addresses the GPU uses into physical memory addresses, and reports its activity through Performance Monitor Counter Groups (PMCGs), which Linux exposes to \code{perf}~\cite{arm-smmuv3}.
The GB10 has \val{smmu.n_total} PMCGs; \val{smmu.n_gpu} belong to the GPU's SMMU, and the rest to the SMMUs of other devices on the chip.
An Arm SMMU consists of translation buffer units (TBUs), which sit beside the devices and cache recent translations, and one translation control unit (TCU), which serves the translations the TBUs miss~\cite{arm-smmuv3}.
The PMCG that responds to copies is the GPU's TCU, and its \code{transaction} event equals \code{pcie\_ats\_trans\_rq}: it counts address translation requests, issued only for host pages.
The bytes per request are not a property of the copy: \val{smmu.bpt_lo} for pinned and \val{smmu.bpt_hi} for registered buffers, \val{smmu.thp}$\times$ fewer requests with 2\,MiB pages, and up to \val{smmu.saw_hi} as one buffer is reused.
GPU to GPU copies, which cross C2C twice, produce none.
A byte calibration obtained with one buffer allocation therefore does not transfer to another, because the count depends on page size and on the state of the translation caches.
The counter remains useful for studying translation, not traffic.

\textbf{L2 counters.}
The GPU's L2 cache, built from slices that NVIDIA calls LTS, has hundreds of hardware counters, and NVIDIA provides three software interfaces to them.
Nsight Compute reads them per kernel by replay and documents their names~\cite{ncu-guide}; Nsight Systems samples a fixed set over time, which on the GB10 excludes memory traffic; and the CUDA Profiling Tools Interface (CUPTI) offers performance monitor (PM) sampling, which reads a chosen subset at fixed intervals while programs run normally, without replay~\cite{cupti-pm}.
We use the third.
Because the GPU has no memory of its own, the bytes it moves are the read misses its L2 fetches and the writes it sends towards memory.
We therefore need L2 counters that count sectors, not requests; that count only reads that missed, since hits never leave the GPU; that count writes; that separate the copy engines from kernels; and that PM sampling accepts in one pass.
Unlike replay, which collects different counters on repeated runs of a kernel, PM sampling configures the counters once and reads them continuously, so it accepts only counters the hardware can collect in a single pass; NVIDIA marks such variants of the L2 counters with the suffix \code{\_realtime}, and CUPTI rejects the others when the sampler is configured.
Table~\ref{tab:cand} lists the candidates we tested and why each was kept or rejected.
Two choices that look right are wrong (Table~\ref{tab:cand}, rows 2 and 3).
The counter of all L2 reads, \nolinkurl{lts__t_sectors_op_read}, includes hits, which are served by the L2 and never reach memory, so it overstates traffic whenever data is reused; \nolinkurl{lts__t_sectors_op_read_lookup_miss} counts only the reads that leave the GPU.
The aperture counters, such as \nolinkurl{lts__t_sectors_aperture_device} and \nolinkurl{lts__t_sectors_aperture_sysmem}, would split traffic into GPU buffers and CPU buffers; Nsight Compute can read them by replay, but PM sampling rejects them, so they cannot be sampled live.

\begin{table}[t]
\centering
\caption{Candidate counters for GPU memory traffic and the test that decided each.}
\label{tab:cand}
\footnotesize
\setlength{\tabcolsep}{3pt}
\arrayrulecolor{blue!30}
\rowcolors{2}{white}{stripe}
\begin{tabular}{|L{0.30\columnwidth}|L{0.13\columnwidth}|L{0.47\columnwidth}|}
\hline
\rowcolor{hdr}
\textcolor{white}{\textbf{Candidate}} & \textcolor{white}{\textbf{Verdict}} & \textcolor{white}{\textbf{Deciding test}} \\ \hline
SMMU PMCG transactions & \no{rejected} & counts translations; \val{smmu.bpt_lo} to \val{smmu.bpt_hi}\,B each, varies by buffer \\ \hline
L2 reads, no \code{lookup\_miss} & \no{rejected} & includes hits; a reread 16\,MiB buffer counts again \\ \hline
L2 aperture variants & \no{rejected} & refused by PM sampling \\ \hline
Nsight Compute DRAM, sysmem & \no{rejected} & kernel replay only; misses driver copies \\ \hline
Five L2 counters (Sec.~\ref{sec:method}) & \yes{kept} & known volumes within \val{ver.gn.maxerr}\% on both systems \\ \hline
\end{tabular}
\end{table}

\section{Measurement method}\label{sec:method}
\textbf{The five counters.}
Three counters measure traffic by source, and two give totals over all sources, so the totals already include the first three and must not be added to them.
Copies are counted by \nolinkurl{lts__t_sectors_srcunit_ce}.
Kernel reads are counted by \nolinkurl{lts__t_sectors_srcunit_tex_op_read_lookup_miss_realtime}, and kernel writes by \nolinkurl{lts__t_sectors_srcunit_tex_op_write_realtime}.
The total is the sum of \nolinkurl{lts__t_sectors_op_read_lookup_miss_realtime} and \nolinkurl{lts__t_sectors_op_write_realtime}.
Each is summed over all L2 slices.

\textbf{Conversion.}
With $S_x$ the sectors counted by counter $x$ in an interval,
\begin{align}
B_{\mathrm{kread}} &= 32\,S_{\mathrm{tex,rd\,miss}}, \qquad B_{\mathrm{kwrite}} = 32\,S_{\mathrm{tex,wr}},\\
B_{\mathrm{copy}} &= 32\,S_{\mathrm{ce}}/2, \qquad\ \ B_{\mathrm{L2}} = 32\,(S_{\mathrm{rd\,miss}} + S_{\mathrm{wr}}).
\end{align}
$B_{\mathrm{L2}}$ is the total GPU memory traffic observed at L2, our estimate of the payload carried over C2C.
The copy engines sit on the GPU die, while both the source and the destination of a copy sit in LPDDR5X, so a copy engine reads each byte across C2C and writes it back across C2C.
The copy counter sees both passes, which is why $B_{\mathrm{copy}}$ divides by two.
Copying 2\,GiB from host to GPU memory therefore reports 2\,GiB as copies and about 4\,GiB of memory traffic, which is the traffic the copy really causes.
The total also includes traffic from units other than the copy engines and the SMs, for example the display engine scanning out the desktop.

\textbf{Sampling from outside.}
A small reader process configures CUPTI PM sampling on the device once, then streams one sample every \val{clk.int_req}\,ms with its interval and five values.
It runs no GPU work and does not attach to the measured program, so a program can be observed without rebuilding or instrumenting it.
It records the GPU sample clock together with the host clock every second; on the GB10 they agree within \val{clk.offset}\,ns (spread \val{clk.spread}\,ns), so samples join with host side power and energy logs.
When the GPU is mostly idle, samples last longer than requested (median \val{clk.int_obs}\,ms); every sample covers its own interval, so totals are unaffected.
To total a phase between two host timestamps, a sample that straddles a boundary is split in proportion to the time it spends on each side.
This assumes the traffic within that sample was uniform, so the fraction of each total that came from split samples is reported with it; for a short phase measured from few samples this fraction shows how much of the result rests on the assumption.
The counters are device wide: they include every process on the GPU.
To remove background traffic, such as the desktop's, we measure the idle GPU's traffic rate just before the work and subtract that rate over the work's duration, as is done for idle power in energy measurement~\cite{epg}:
\begin{equation}
B_{\mathrm{net}} = B_{\mathrm{run}} - (B_{\mathrm{idle}}/t_{\mathrm{idle}})\,t_{\mathrm{run}}.
\end{equation}
This assumes the background stays at its idle rate, so $B_{\mathrm{net}}$ is an estimate, and raw values are always kept.

\textbf{Systems.}
We used an \val{gn.product} (Linux \val{gn.kernel}) and an \val{sp.product} (Linux \val{sp.kernel}), both with driver \val{gn.driver}, CUDA \val{gn.cuda} and Nsight Compute \val{plat.ncu}.
The GN100 was dedicated to the measurements, with a desktop running; the DGX Spark is a shared machine without a desktop.

\textbf{Validation design.}
A test program moves an exact volume with CUDA's own operations: \code{cudaMemcpy} from host to GPU and back (copy engines), \code{cudaMemset} (kernel writes), and a GPU to GPU copy, which the driver performs with a kernel that reads and writes.
For each, the reader measures the window in which the work ran, subtracts the background, and compares with the known volume; each check runs three times and reports the middle value.

\section{Results}
\begin{figure}[t]
\centering
\includegraphics[width=\columnwidth]{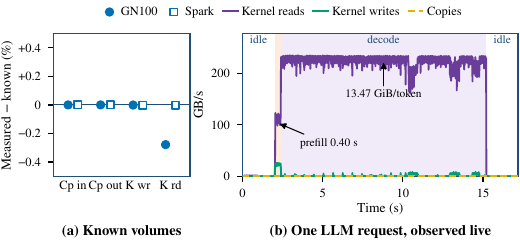}
\caption{(a) Error of the measured against the known volume on both systems: copies into (Cp in) and out of (Cp out) GPU memory, 2\,GiB each; kernel writes (K wr) and reads (K rd), 4\,GiB each.
(b) One request to a vLLM server (Mistral 7B, \val{llm.prompt_tokens} prompt tokens, \val{llm.gen_tokens} generated): prefill reads the weights about \val{llm.prefill_passes} times in \val{llm.prefill_s}\,s; decode reads \val{llm.decode_gib_tok}\,GiB per token, while its writes and copies stay near zero.}
\label{fig:results}
\end{figure}

\textbf{Known volumes.}
Measured and known volumes agree within \val{ver.gn.maxerr}\% on the GN100 and within 0.005\% on the DGX Spark, the rounding limit of our ratios (Fig.~\ref{fig:results}a).
The DGX Spark runs no desktop, so during a check nothing but the test program moves memory; the GN100's larger error comes from its desktop, as the read check shows.
Copies in both directions and kernel writes measure \val{ver.gn.h2d}, \val{ver.gn.d2h} and \val{ver.gn.write} of the known volume on the GN100 and \val{ver.sp.h2d}, \val{ver.sp.d2h} and \val{ver.sp.write} on the DGX Spark; kernel reads measure \val{ver.gn.read} and \val{ver.sp.read}.
Because the desktop's screen updates read memory, the read check is the hardest on the GN100, and its error of \val{ver.gn.read.err}\% is the largest of all checks; without background subtraction its error would be larger, and the small negative error that remains shows that the desktop's bursty background is subtracted only approximately.
Three further properties support reading the counts as traffic.
First, they track the volume moved, not the number of copies: 2\,GiB copied as 8\,$\times$\,256, 4\,$\times$\,512 or 16\,$\times$\,128\,MiB gives \val{split.lo} to \val{split.hi}\,GiB, the excess being the program's \val{split.startup}\,MiB of CUDA start up copies.
Second, they exclude reuse: rereading a \val{reuse.mib}\,MiB buffer, which fits in L2, gives read misses of \val{reuse.ratio} of the bytes read.
Third, the total follows the path: on the GN100, a whole run that copies 4\,GiB from one GPU buffer to another gives \val{d2d.c2c}\,GiB of memory traffic, 4\,GiB read and 4\,GiB written by the copy, 0.5\,GiB written by the program clearing its buffers, and the rest from the desktop and CUDA start up; on a discrete GPU, none of it would cross the host link.
Copies issued by PyTorch, observed from outside, measure \val{pt.measured}\,GiB for \val{pt.known}\,GiB.

Two properties of the L2 counts, found with Nsight Compute, also bear on their use.
They count distinct 32\,B sectors, not requested bytes: in all tested access patterns the count stayed within \val{sp.n7.max_pct}\% of the sectors touched.
Writes to pageable memory produced \val{gn.e7.malloc_w}\% more write sectors than expected, with about \val{gn.e7.malloc_extra_mb}\,MB of extra read sectors per 256\,MiB pass, on both systems; we have not established whether this is extra traffic or a counting effect, so traffic should be compared only between workloads that use the same allocation type.
The separate reader does not measurably slow the program: timed \val{ovh.runs} times with and without the reader, a GPU job moving 8\,GiB changed its median time by at most \val{ovh.max}\% on the two systems, within the spread between repeated runs (at least \val{ovh.spread}\%).
The released tool passed all 19 of its tests on both systems, covering installation, sampling, known volumes, phases, LLM observation and clean removal; each is listed with its command and result in the artifact's test matrix~\cite{pravaha}.

\textbf{An LLM server observed from outside.}
Fig.~\ref{fig:results}b shows one request to a vLLM server running Mistral 7B (\val{mis.weights}\,GiB of weights), measured without changing the server.
During decode the GPU reads \val{llm.decode_gib_tok}\,GiB per token, moves \val{llm.decode_gbs}\,GB/s of reads plus writes, and generates \val{llm.decode_tok_s} tokens/s: each token reads all weights except the unused rows of the embedding table, plus the KV cache of the \val{llm.prompt_tokens} earlier tokens, \val{mis.kv_kib}\,KiB each (Table~\ref{tab:llm}).
Prefill reads \val{llm.prefill_kread}\,GiB in \val{llm.prefill_s}\,s, about \val{llm.prefill_passes} passes over the weights for a prompt of \val{llm.prompt_tokens} tokens, consistent with processing the prompt in chunks, and writes \val{llm.prefill_kwrite}\,GiB.
Copies are zero throughout: after loading, inference moves its data through kernels only.
The decode rate is about 83\% of the 273\,GB/s that NVIDIA specifies for the GB10's memory~\cite{spark-hw}, consistent with decode being constrained by memory bandwidth on this system.
Table~\ref{tab:llm} also adds a second model, Qwen3 4B, and sets each measurement against what the model sizes predict.
The expectation ignores activations and small tensors, so we read the agreement as evidence that decode streams the weights and the KV cache, not as an exact test.
Prefill behaves differently: its reads and writes together run at \val{llm.prefill_gbs}\,GB/s, about half the rate of decode, yet it reads the weights several times, because each chunk of the prompt reuses them across many tokens and spends more of its time computing.
Its \val{llm.prefill_kwrite}\,GiB of writes are mostly the KV cache and activations of the prompt; decode writes little.
Between requests the server is not silent; its idle traffic, measured before the request, is removed by the baseline.

\begin{table}[t]
\centering
\caption{Decode traffic per token: expected from model sizes and measured from outside the server.}
\label{tab:llm}
\footnotesize
\setlength{\tabcolsep}{4pt}
\arrayrulecolor{blue!30}
\rowcolors{2}{white}{stripe}
\begin{tabular}{|L{0.28\columnwidth}|L{0.44\columnwidth}|r|}
\hline
\rowcolor{hdr}
\textcolor{white}{\textbf{Case}} & \textcolor{white}{\textbf{Expected}} & \textcolor{white}{\textbf{Measured}} \\ \hline
Qwen3 4B, short prompt & all weights: \val{qwen.weights} & \val{qwen.gib_tok} \\ \hline
Mistral 7B, short prompt & all weights except the unused rows of the embedding table: \val{mis.short_expect} & \val{mis.short_gib_tok} \\ \hline
Mistral 7B, \val{llm.prompt_tokens} prompt tokens & row above plus the KV cache of \val{llm.prompt_tokens} tokens at \val{mis.kv_kib}\,KiB each: \val{mis.long_expect} & \val{llm.decode_gib_tok} \\ \hline
\end{tabular}
\\[2pt]{\scriptsize All values in GiB per generated token. Expected values ignore activations and small tensors.}
\end{table}

\section{Discussion, Limitations, and Conclusion}\label{sec:limits}
\textbf{Summary of findings.}
On the GB10, GPU memory traffic is visible only from the GPU side, at its L2.
Without these counters, the bytes that kernels move, which are nearly all of the traffic of LLM inference, cannot be measured live; with them, an unmodified server shows decode streaming its weights and KV cache once per token, near the memory's specified limit.
Studies of unified CPU and GPU memory on Grace Hopper~\cite{schieffer2024gh} and MI300A~\cite{wahlgren2025mi300a} rely on counters the GB10 does not expose, and GB10 studies measure traffic per kernel with Nsight Compute~\cite{zhu2026sawtooth}.
CUPTI PM sampling was designed to run inside the profiled program~\cite{cupti-pm}, and SMMU counters monitor translation~\cite{arm-smmuv3}; our contribution is to establish which counters track traffic and to sample them from a separate process.

\textbf{Limitations.}
The counters measure traffic observed at L2, not bytes on the C2C wire, which would include protocol overhead and translation traffic; reading them as C2C payload relies on where NVIDIA places the memory controllers~\cite{hotchips-gb10}, with which our known volume checks agree.
The L2 is a write back cache, so data written repeatedly before eviction may be counted more often than it reaches memory, a case our checks do not test.
The counts are device wide, exclude CPU memory traffic, and allow one reader at a time; and we validated two systems with one driver.

\textbf{Conclusion.}
The GB10 exposes no C2C traffic counter, yet five L2 counters, sampled from a separate process, measure its GPU memory traffic live, split into copies and kernel reads and writes.
Joined with power samples on the same host clock, they open the energy per byte of GPU memory traffic to measurement~\cite{epg}, and the study of how inference slows CPU work in the shared memory.
The tool and data are released as open source~\cite{pravaha}.

\IfFileExists{IEEEtran.bst}{\bibliographystyle{IEEEtran}}{\bibliographystyle{plain}}
\bibliography{refs}

\end{document}

%% file: figures/fig1_system.tex
\definecolor{fcopy}{HTML}{E8B000}
\definecolor{fread}{HTML}{6A3D9A}
\definecolor{fwrite}{HTML}{009E73}
\definecolor{fmeas}{HTML}{0072B2}
\definecolor{bgpu}{HTML}{E6F4EA}
\definecolor{bcpu}{HTML}{E4EEF8}
\definecolor{bl2}{HTML}{FFF3B0}
\definecolor{bsm}{HTML}{EFE6F7}
\definecolor{bce}{HTML}{FFF0B8}
\definecolor{bc2c}{HTML}{FDEBDD}
\definecolor{bmem}{HTML}{D9F0EE}
\providecommand{\fsub}[1]{{\fontsize{5.8}{6.6}\selectfont #1}}
\begin{tikzpicture}[
  font=\fontsize{7}{8}\selectfont,
  >={Stealth[length=1.5mm,width=1.15mm]},
  die/.style={line width=0.6pt, fill=white, rounded corners=1.5pt},
  blk/.style={draw=blue!45!black, line width=0.45pt, fill=white, rounded corners=1pt, align=center, inner sep=1pt},
  sw/.style={draw=fmeas, line width=0.6pt, dash pattern=on 2pt off 1pt, fill=white, rounded corners=2pt,
             align=center, inner sep=1pt},
  tag/.style={draw=fmeas, line width=0.4pt, fill=fmeas!8, rounded corners=0.8pt, inner sep=0.8pt,
              font=\fontsize{5.6}{6}\selectfont, text=fmeas!80!black},
  title/.style={font=\fontsize{7}{8}\selectfont\bfseries, anchor=north west, inner sep=1.5pt},
  cp/.style={->, line width=1.15pt, color=fcopy},
  rd/.style={->, line width=0.95pt, color=fread},
  wr/.style={->, line width=0.95pt, color=fwrite},
  ms/.style={->, line width=0.75pt, color=fmeas, dash pattern=on 2.2pt off 1.2pt},
]
\node[die, draw=green!45!black, minimum width=7.6cm, minimum height=2.15cm, anchor=south west] (gpu) at (0,3.05) {};
\fill[bgpu, rounded corners=1.5pt] ([shift={(0.25pt,-0.25pt)}]gpu.north west) rectangle ([shift={(-0.25pt,0.25pt)}]gpu.south east);
\node[title] at (gpu.north west) {GPU die (Blackwell): no memory of its own};
\node[blk, fill=bsm, draw=fread!80!black, minimum width=2.55cm, minimum height=0.56cm, anchor=south west] (sm) at (0.2,4.25)
     {SMs \fsub{(kernels: GPU code)}};
\node[blk, fill=bce, draw=fcopy!85!black, minimum width=2.55cm, minimum height=0.56cm, anchor=south east] (ce) at (7.4,4.25)
     {Copy engines \fsub{(cudaMemcpy)}};
\node[blk, fill=bl2, draw=black!80, line width=0.8pt, minimum width=7.2cm, minimum height=0.72cm, anchor=south west] (l2) at (0.2,3.2) {};
\node[anchor=north, inner sep=1.5pt, font=\fontsize{6.6}{7.6}\selectfont] at (l2.north) {\textbf{L2 cache} (LTS slices): every request to memory passes it; \textbf{counted here}};
\node[tag, anchor=south] at (1.0,3.24) {tex read miss};
\node[tag, anchor=south] at (2.25,3.24) {tex write};
\node[tag, anchor=south] at (6.12,3.24) {ce};
\fill[bc2c] (0,2.45) rectangle (7.6,3.05);
\draw[orange!55!black, line width=0.4pt] (0,2.45) -- (7.6,2.45) (0,3.05) -- (7.6,3.05);
\node[font=\fontsize{7}{8}\selectfont\bfseries, fill=bc2c, inner sep=0.5pt, text=orange!40!black] at (3.95,2.86) {NVLink C2C};
\node[fill=bc2c, inner sep=0.5pt, text=orange!35!black] at (3.95,2.57) {\fsub{no traffic counter exposed}};
\node[die, draw=blue!55!black, minimum width=7.6cm, minimum height=2.0cm, anchor=south west] (cpu) at (0,0.45) {};
\fill[bcpu, rounded corners=1.5pt] ([shift={(0.25pt,-0.25pt)}]cpu.north west) rectangle ([shift={(-0.25pt,0.25pt)}]cpu.south east);
\node[blk, minimum width=7.2cm, minimum height=0.5cm, anchor=north west] (mc) at (0.2,2.3)
     {memory controllers and system level cache};
\node[title, anchor=south west] at (cpu.south west) {CPU die (Arm)};
\node[blk, minimum width=1.65cm, minimum height=0.5cm, anchor=south west] (cores) at (2.3,0.6) {Arm cores};
\node[blk, minimum width=1.85cm, minimum height=0.5cm, anchor=south west] (smmu) at (4.5,0.6)
     {SMMU \fsub{(translation)}};
\node[draw=fmeas, dash pattern=on 2pt off 1pt, line width=0.5pt, fill=fmeas!6, rounded corners=1.5pt, minimum width=7.6cm, minimum height=0.6cm, anchor=north west] (swband) at (0,-0.12) {};
\node[anchor=west, inner sep=1.5pt, font=\fontsize{7}{8}\selectfont\bfseries, text=fmeas!70!black] at (0.08,-0.42) {Software (ours)};
\node[blk, fill=fmeas, text=white, draw=fmeas!70!black, minimum width=4.6cm, minimum height=0.42cm, anchor=east] (rdr) at (7.4,-0.42)
     {\textbf{pravaha-reader} \fsub{a process on the Arm cores}};
\node[blk, fill=bmem, draw=teal!70!black, minimum width=7.6cm, minimum height=0.42cm, anchor=north west] (mem) at (0,0.28)
     {\textbf{LPDDR5X} (128\,GB), shared by CPU and GPU};
\draw[teal!70!black, line width=1.6pt] (4.22,1.8) -- (4.22,0.28);
\draw[rd] (1.0,2.3) -- (1.0,3.2);
\draw[rd] (1.0,3.92) -- (1.0,4.25);
\draw[wr] (2.25,4.25) -- (2.25,3.92);
\draw[wr] (2.25,3.2) -- (2.25,2.3);
\draw[cp] (6.12,2.3) -- (6.12,3.2);
\draw[cp] (6.12,3.92) -- (6.12,4.25);
\draw[cp] (6.95,4.25) -- (6.95,3.92);
\draw[cp] (6.95,3.2) -- (6.95,2.3);
\draw[ms] (7.4,3.56) -- (7.82,3.56) -- (7.82,-0.42) -- (rdr.east);
\node[rotate=90, text=fmeas!85!black, fill=white, inner sep=0.8pt] at (7.82,2.15) {\fsub{CUPTI PM sampling, 10\,ms}};
\begin{scope}[shift={(0.05,-1.0)}]
  \draw[cp] (0,0) -- (0.35,0);      \node[anchor=west, inner sep=1pt] at (0.37,0) {copy path};
  \draw[rd] (1.55,0) -- (1.9,0);    \node[anchor=west, inner sep=1pt] at (1.92,0) {kernel reads (misses)};
  \draw[wr] (4.3,0) -- (4.65,0);    \node[anchor=west, inner sep=1pt] at (4.67,0) {kernel writes};
  \draw[ms] (6.3,0) -- (6.65,0);    \node[anchor=west, inner sep=1pt] at (6.67,0) {counting};
\end{scope}
\end{tikzpicture}